\documentclass[preprint,showpacs,showkeys,preprintnumbers,amsmath,amssymb,amsfonts]{revtex4}

\usepackage{graphicx}
\usepackage{epsfig}
\usepackage{bm}
\usepackage{verbatim}

\begin{document}

\title{Molecular mean field theory for liquid water}
\author{Jampa Maruthi Pradeep Kanth}
\email{jmpkanth@imsc.res.in}
\author{Ramesh Anishetty} 
\email{ramesha@imsc.res.in}
\affiliation{The Institute of Mathematical Sciences, C.I.T. Campus, Tharamani, Chennai  600113,  India}
\date{\today}

\begin{abstract}
Attractive bonding interactions between molecules typically have inherent conservation laws which influence the statistical properties of such systems in terms of corresponding sum rules. We considered lattice water as an example and enunciated the consequences of the sum rule through a general computational procedure called ``Molecular mean field'' theory. Fluctuations about mean field are computed and many of the liquid properties have been deduced and compared with Monte Carlo simulation, molecular dynamics and experimental results. Large correlation lengths are seen to be a consequence of the sum rule in liquid phase. Long range Coulomb interactions are shown to have minor effects on our results.
\end{abstract}


\maketitle

\section{Introduction} \label{introduction}

Molecules in bulk interact with each other consistent with the underlying quantum mechanical dynamics, for example, hydrogen bond, variety of sulphur bonds etc. These interactions, though quite strongly attractive compared to ambient thermal energy, still fluctuate only when the concerned molecules are very close to each other (of the order of Angstrom). These molecular interactions are countable and categorizable. For concreteness, consider water molecules. Each water molecule has two hydrogen arms and two lone pair arms. Interaction between water molecules is via these oppositely polarized bond arms and when a hydrogen arm of one molecule bonds with lone pair of a neighboring molecule, it results in ``hydrogen bond". So, for any number of water molecules ($\rho$) the number of hydrogen bonds (HB) and number of dangling bonds (DB) i.e., lone-pair and hydrogen arms which are not hydrogen-bonded, satisfy a ``sum rule". It states that the arms of water molecule are either hydrogen-bonded or not i.e.,
	\begin{equation}
		\text{DB} + 2 \ \text{HB} = 4 \rho
	\label{sumrule}	
	\end{equation}
 
Now, if we consider a bulk of water molecules the above equation still holds when DB, HB and $\rho$ are appropriately defined per unit volume. In other words, the local topology of molecular interactions implies a sum rule which is also true in the bulk for any thermodynamic conditions such as temperature, pressure, volume. Furthermore, this is also independent of other interactions in the dynamical system such as van der Waals' (vdW), Coulombic etc. These facts are not surprising since Eq.(\ref{sumrule}) is a topological constraint which is insensitive to many details of the dynamics.

In most descriptions of solid phase (ice or glasses), by design, the above sum rule is maintained explicitly. In the gas phase where the density of hydrogen bonds is negligible the sum rule is again trivially satisfied. If there are significant HB then the system can be analyzed by introducing appropriate density of dimers which still satisfy the sum rule thermodynamically.

Liquid phase is typically studied using molecular dynamics (MD) and Monte Carlo (MC) simulations of popular models \cite{RahmanJCP1971Vol55, StillingerJCP1974Vol60, BerendsenJPC1987Vol91, KusalikScience1994Vol265, JorgensenJCP1983Vol79, JorgensenMolPhy1985Vol56, MahoneyJCP2000Vol112, WallqvistRCC1999, JorgensenPNAS2005Vol102} where in again the sum rule is maintained at every step of the dynamics and many bulk and other correlation properties \cite{StillingerScience1980Vol209} are well reproduced and are consistent with experiments \cite{LevyDFS1967, SoperJCP1994Vol101, SorensonJCP2000Vol113, WernetScience2005}.

Liquid phase is also described phenomenologically by a mean field which invariably is taken to be density field and fluctuations there of are then postulated to obey Ornstein-Zernike (OZ) equation \cite{HansenSimpleLiquids3ed, BarkerRMP1976,ChandlerARPC1978}. Furthermore, all molecular interactions have the property that they cannot approach each other closer than certain minimum distance, usually incorporated as hard sphere repulsion in the interaction potential between the molecules. This in the liquid phase gives us the well known vdW term in the equation of state, often referred to as density saturation effect \cite{HansenSimpleLiquids3ed}. Phenomenologically, such saturation effects can be incorporated in the OZ equation and thereby understand important components of density fluctuations namely coordination shell peaks using approximations like Percus-Yevick, hypernetted-chain etc \cite{HansenSimpleLiquids3ed}. But these descriptions are not as successful as MD simulations, primarily because by design in the construction of phenemonological models one does not envisage fluctuations in HB or DB densities. Consequently, they cannot satisfy the sum rule in large subvolumes where surface effects are negligible. The sum rule argument suggests that to describe water we need atleast two fluctuating fields to be self-consistent. 

Some important developments in the past on systems with hydrogen-bond' like interactions are discussed below. Wertheim's work on statistical thermodynamics of fluids with directional interactions consists of systematic graph-theoretic expansions of n-mer tree-like configurations and provides OZ-like operator equation in terms of total density and singlet-density fields of the system \cite{Wertheim1JStatPhy1984Vol35, Wertheim2JStatPhy1984Vol35, Wertheim1JStatPhy1986Vol42}. Another significant work on associating fluids was by Andersen who provided a Mayer-like density expansion technique to deal with directional interactions \cite{AndersenJCP1973Vol59}. Both the descriptions work very well at high temperatures and also at densities corresponding to liquid-vapor coexistence. An extension to Wertheim's theory was proposed as SAFT \cite{ChapmanFPhEq1989, Muller2001} wherein the equation of state is phenomenologically extended to include contributions from complex geometries of constituent molecules which interact through directional forces. Another phenomenological description aimed at describing mainly the low temperature anamolies in fluids with characterstic orientation-dependent interactions was given by Truskett et al \cite{TruskettJCP1999Vol111}. Here in again, the equation of state is approximated with vdW like terms which are then phenomenologically supplemented with hydrogen-bonding contribution that captures certain high-density features.

In this paper we propose to study a simple model Hamiltonian which incorporates the molecular hydrogen-bonding interactions. To accomodate the hard sphere repulsion we envisage the system on a hypercubic lattice and the model is essentially a slight generalization of the Pauling's model of water \cite{PaulingJACS1935}. The partition function corresponding to the lattice model is analyzed by introducing appropriate discrete lattice fields. It is shown that the sum rule is automatically true in the bulk. `Molecular mean field' (MMF) approximation extremizes the partition function in terms of the defined auxiliary fields. In addition, all the observables such as $\rho$, HB and DB densities are functionals of these auxiliary fields. One of the mean field equation which implies sum rule also implicates the ``equation of network" i.e., a relation between equilibrium densities of HB and $\rho$. We study the equation of state and various mean field fluctuations in terms of correlation functions. We also considered long range Coulomb interaction and studied its consequences. Subsequently, an MC simulation study of the model is pursued and compared with the mean field theory quantitatively. We also discussed results of our analysis in the context of experiments and MD simulations. A brief discussion on future directions is presented at the end.

\section{Model for water} \label{model}

\begin{figure*}[h]
		\epsfig{file= 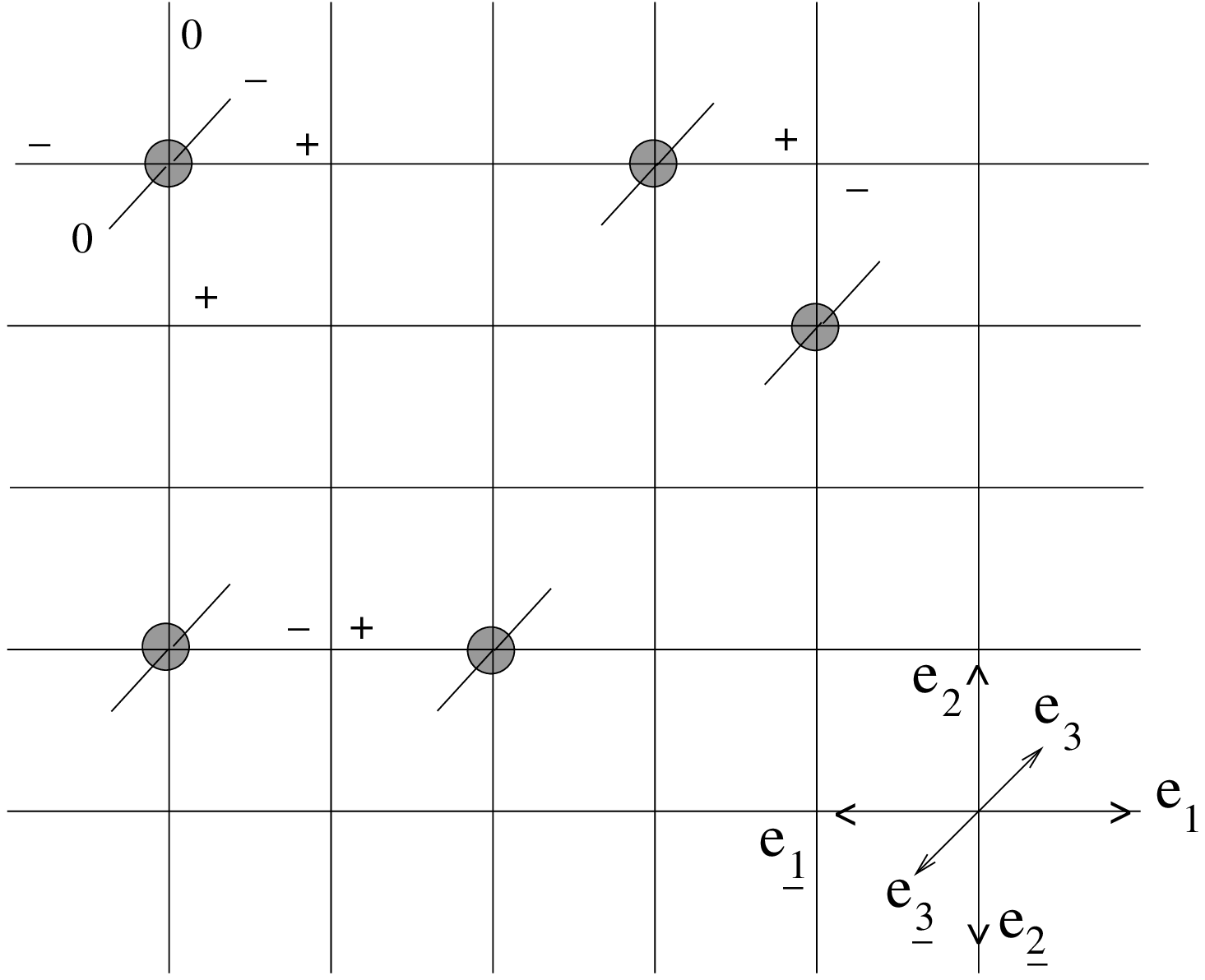 ,width=0.5\linewidth,height=0.3\textheight}
		\caption{\label{allowedconfigs} Allowed configurations : (grey circle) water site, (+) H arms, (-) L arms and (0) no arms; two types of dimer hydrogen-bonded via (+) H arm and (-) L arm; (right bottom corner)  unit vectors' definition on cubic lattice.}
	\end{figure*}

On a three dimensional hypercubic lattice we define $W(x) = \lbrace 0, 1 \rbrace$ corresponding to water being absent or present respectively, at the site $x$. At each occupied site we define six arms $H_{\alpha}(x) = \lbrace 0, \pm 1 \rbrace$, where $\alpha = \lbrace \pm 1, \pm 2, \pm 3 \rbrace$ corresponding to six directions around the site. $H_{\alpha}(x) = 0$  corresponds to no arm on the corresponding link, $+1$ to that of hydrogen arm (H), $-1$ for lone pair arm (L). The constraints between $W(x)$ and $H_{\alpha}(x)$ being,
	\begin{eqnarray}
		\sum_{\alpha} H_{\alpha}^{2}(x) &=& 4 W(x) \label{H2constraint}
		\\ \sum_{\alpha} H_{\alpha}(x)  &=& 0 \label{Hconstraint}
	\end{eqnarray}
which imply that every water has two H arms and two L arms only. A hydrogen bond is realized when two water molecules two lattice units apart have their H and L arms meet at a site, as shown in Fig.(\ref{allowedconfigs}).

\begin{figure*}[h]
		\epsfig{file= 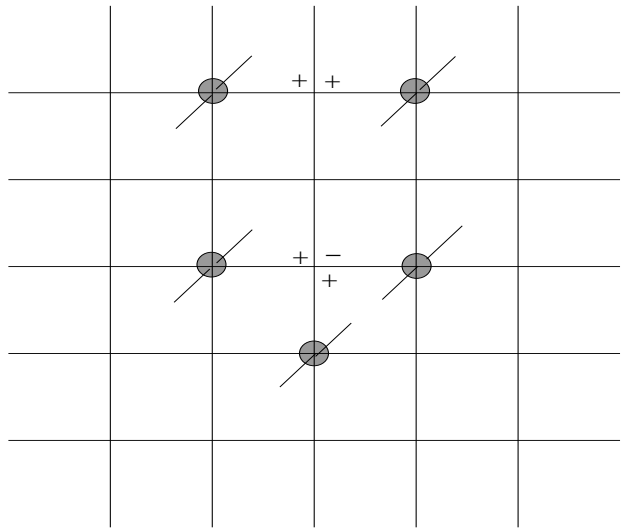 ,width=0.5\linewidth,height=0.3\textheight}
		\caption{\label{disallowedconfigs} Disallowed configurations : (+) H arms (or (-) L arms) of two neighboring molecules meeting at a site, more than two non-zero arms meeting at a site.}
	\end{figure*}

At any site and on any link on the lattice the possible configurations are as shown in Fig. (\ref{allowedconfigs}). We assign an energy $-\tilde{\nu}$ for every unpaired arm (H or L), and $- \tilde{\lambda}$ for every hydrogen bond. We disallow for any two H (or L) arms to meet at the same site; also, more than two arms are disallowed to meet at a site as in Fig.(\ref{disallowedconfigs}).

To implement the constraints as shown in Fig.(\ref{disallowedconfigs}) in our analysis it is useful to define two discrete integer fields $b(x)$, $q(x)$:
	\begin{subequations}
	\label{bqdefinition}
	\begin{eqnarray}
		b(x) &=& \sum_{\alpha} H_{\alpha}^{2}(x+{e}_{\alpha})
		\\ q(x) &=& \sum_{\alpha} H_{\alpha}(x+{e}_{\alpha})
	\end{eqnarray}
	\end{subequations}
where ${e}_{\alpha}$ is a lattice unit vector in the direction $\alpha$ with the property ${e}_{\alpha} = -{e}_{\underline{\alpha}}$.  The discrete field $b(x)$ counts the number of non-zero arms (approaching arms) in the neighborhood of site $x$, while $q(x)$ measures the charge i.e, the difference between number of H arms and L arms meeting at site $x$. By construction, $b(x)$ varies between $0 \ \text{and} \ 6$ on a three dimensional hyper-cubic lattice and by imposing the condition that $b(x) \leq 2$ in our analysis we ensured that no more than two arms can meet at a site. $q(x)$ in turn varies between $-b(x)$ to $b(x)$. In terms of these variables the Hamiltonian is :
	\begin{equation}
		H = \sum_{x} \left( - \tilde{\nu} \delta({b(x),1}) - \tilde{\lambda} \delta({b(x),2}) \delta({q(x),0}) \right) \label{Hamiltonianredefinition}
	\end{equation}
with constraints, 
	\begin{subequations}
	\label{bqconstraints}
	\begin{eqnarray}
		 && W(x) b(x)  =  0
		 \\ && 0 \leq b(x) \leq 2
		 \\ && q(x)  = \lbrace -b(x), -b(x)+2, \ldots, b(x)-2, b(x) \rbrace
	\end{eqnarray}
	\end{subequations}
The range of $q(x)$ follows from Eq.(\ref{bqdefinition}). The Kronecker delta function denoted here as $\delta({p,q})$ is defined as $\delta({p,q}) = 1$ for $p = q$ and $0$ otherwise.
	
We now write the grand canonical partition function for our system at a finite chemical potential $\tilde{\mu}$ for water and inverse temperature $\beta$:
	\begin{widetext}
	\begin{subequations}
	\label{Zequation}
	\begin{eqnarray}
		Z &=&  \nonumber \left[ \prod_{x}  \sum^{'}_{W(x), H_{\alpha}(x) \atop b(x), q(x)} \frac{1}{(2N+1)^2} \sum_{\eta(x), \phi(x)} \right] \text{exp}  \sum_{x} \left[{\displaystyle \ \beta \tilde{\nu} \ \delta({b(x),1}) + \beta \tilde{\lambda} \ \delta({b(x),2}) \ \delta({q(x),0}) + \beta \tilde{\mu} W(x)} \right. 
		\\&&  \qquad + \left. {i \frac{\pi}{N}} {\displaystyle \eta(x) \left\lbrace \sum_{\alpha} H^{2}_{\alpha}(x + {e}_{\alpha}) - b(x) \right\rbrace}  + {i \frac{\pi}{N}} \phi(x) \left\lbrace \sum_{\alpha} H_{\alpha}(x + {e}_{\alpha}) - q(x) \right\rbrace \right] 
 		\\ & \equiv & \left[\displaystyle  \prod_{x} \frac{1}{(2N+1)^2}\sum_{\eta(x), \phi(x)} \right] \prod_{x} Z_{site}(x)
	\end{eqnarray}
	\end{subequations}
	\end{widetext}
where we have introduced two auxiliary fields $\eta(x)$ and $\phi(x)$ to impose the constraints Eq.(\ref{bqdefinition}). The discrete fields $\eta$ and $\phi$ take integer values in the range $[-N, N]$ at every site, where $N$ is any suitable large integer (greater than $8$). The prime over summation refers to sum being restricted to constraints Eqs.(\ref{H2constraint}, \ref{Hconstraint}, \ref{bqconstraints}). The introduction of auxiliary fields $\eta(x)$ and $\phi(x)$ allows summation over other discrete fields $W(x), H_{\alpha}(x), b(x), q(x)$ within their respective allowed range at each site $x$ without any restrictions from the neighborhood configurations i.e., as if a single site functional $Z_{site}$.

The $Z_{site}$ expression thus obtained is stated below. For brevity in the following expression  $\frac{ \pi}{N} \eta$ is written as $\eta$, $\frac{\pi}{N} \phi$ as $\phi$ and further using fugacity definitions $\nu \equiv \exp(\beta \tilde{\nu})$, $\lambda \equiv \exp(\beta \tilde{\lambda})$ and $\mu \equiv \exp(\beta \tilde{\mu})$, 
	\begin{widetext}
	\begin{eqnarray}
		Z_{site} &=&  1 + 2 {\nu}e^{\displaystyle \ -i \eta(x)} \text{cos}\phi(x)  + {\lambda} e^{\displaystyle \ -2i \eta(x)} \label{Zsiteequation}
		\\ \nonumber && + {\mu} \left\lbrace  \text{exp} \left[{\ i \eta(x + {e}_{1}) + i \eta(x + {e}_{\underline{1}}) + i \eta(x + {e}_{2}) + i \eta(x + {e}_{3})} \right]  \right.
		\\  \nonumber && \qquad \qquad  \left.  \left( 2 \text{cos} ( \nabla_{1}\phi + \nabla_{\underline{1}}\phi - \nabla_{2}\phi - \nabla_{3}\phi )  + 4 \text{cos}(\nabla_{1}\phi - \nabla_{\underline{1}}\phi) \text{cos}(\nabla_{2}\phi - \nabla_{3}\phi)  \right)   \right.
		\\  \nonumber && \left. \qquad + \ldots \text{other orientations} \right\rbrace
	\end{eqnarray}
	\end{widetext}

\begin{figure*}[h]
		\epsfig{file=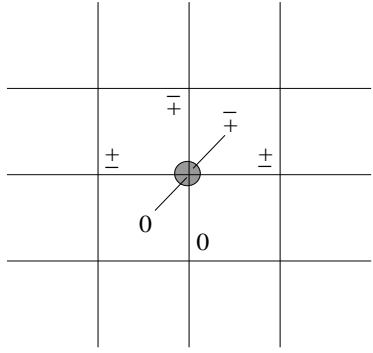 ,width=0.3\linewidth,height=0.2\textheight}
		\caption{\label{fig_zsite} A set of orientations consistent with Eqs.(\ref{H2constraint},\ref{Hconstraint}) and corresponding to representative weight in $\mu$ term of Eq.(\ref{Zsiteequation}). }
	\end{figure*}

Various terms in Eq.(\ref{Zsiteequation}) follow from the fact that at any site $x$ there are only following contributions to the partition function : (i) unity for vacuum, (ii) $\nu$ term for unpaired H or L arms (dangling bond), (iii) $\lambda$ term for hydrogen bond and (iv) $\mu$ term for water with all its possible orientations suitably weighted. For the $\mu$ term there are $\left[ 6 \atop 4 \right]$ arm orientations of having four non-zero arms around any water site and each of these in turn has $\left[ 4 \atop 2 \right]$ possible charge orientations corresponding to Fig.(\ref{fig_zsite}). Explicitly in the Eq.(\ref{Zsiteequation}) we have summed up these $\left[4 \atop 2\right]$ charge orientations. Similarly other arm orientations are implicated in the Eq.(\ref{Zsiteequation}). It is useful to note the following identity in implementing this computation.
	\begin{eqnarray}
		\displaystyle \sum_{x, \alpha}\phi(x) H_{\alpha}(x + {e}_{\alpha})  =  \sum_{x,\alpha}  \phi(x+{e}_{\alpha}) H_{\alpha}(x)  =   \sum_{x,\alpha}  \phi(x) \nabla_{\alpha} H_{\alpha}(x) \label{phiidentity}
	\end{eqnarray}
where $\nabla_{\alpha} f(x) \equiv (f(x+{e}_{\alpha}) - f(x))$. The last equation follows from the constraint Eq.(\ref{Hconstraint}).
	
The observables in the system such as water density $\rho$, hydrogen bond density (HB) and dangling bond (DB) are calculated as follows :
	\begin{subequations}
	\label{observables}
	\begin{eqnarray}
		\rho  & = & \mu \frac{\partial}{\partial {\mu}} \left(\frac{1}{V} \ln Z \right)
		\\ \text{HB}  & = & \lambda \frac{\partial}{\partial {\lambda}} \left(\frac{1}{V} \ln Z \right)
		\\ \text{DB} & = & \nu \frac{\partial}{\partial {\nu}} \left(\frac{1}{V} \ln Z \right)
	\end{eqnarray}
	\end{subequations}
where $V$ is the volume i.e., total number of lattice points.

$\eta(x)$ and $\phi(x)$ are discrete fields varying in the range $[-N,N]$. By construction the partition function is independent of $N$ for $N \geq 8$. In practice it is convenient to evaluate this partition function by taking $N \rightarrow \infty$, whereupon effective $\eta(x)$ and $\phi(x)$ become continuous fields. We implement this limiting procedure and check if the sum rule is obeyed. In the $N \rightarrow \infty $ limit, summation over $\eta$ and $\phi$ is replaced by integrals. The resulting functional integral has the following trivial property :
	\begin{equation}
		\displaystyle \left[ \prod_{x} \int \frac{d\eta(x)}{2 \pi} \ \frac{d\phi(x)}{2 \pi} \right] \sum_{y} \frac{d}{d\eta(y)} \prod Z_{site} = 0 \label{etaderivative}
	\end{equation}
	
Taking derivatives explicitly in the above gives terms proportional to $\nu$, $\lambda$, $\mu$. Since these are being summed at all sites each of these terms can be regrouped in terms of derivatives of $\nu$, $\lambda$, and $\mu$ as :
	\begin{equation}		
		i \left[ - \nu \frac{\displaystyle \partial}{\displaystyle \partial \nu} - 2 \lambda \frac{\displaystyle \partial}{\displaystyle \partial \lambda} + 4 \mu \frac{\displaystyle \partial}{\displaystyle \partial \mu} \right] Z  =  0 \label{fugacityderivative}
	\end{equation}

The $\mu$ dependent term in the Eq.(\ref{fugacityderivative}) has contributions from the four neighboring sites. Since all sites are being summed over, the $\mu$ term in Eq.(\ref{etaderivative}) gets each contribution four times. We notice that this equation is precisely the ``sum rule" constraint Eq.(\ref{sumrule}). This demonstrates that the sum rule in terms of continuous auxiliary fields is automatically true.

\section{Molecular mean field theory} \label{MMF}

Next, we evaluate the functional integral within the MMF approximation. The partition function's integrand can be envisaged as a product of field dependent phase factors at each site. When we enumerate them site-by-site the phase factors cancel exactly corresponding to physically allowed configurations. Evaluating along this procedure amounts to the standard high temperature or Mayer-like expansion. Instead we attempt an approximate method wherein we first notice that if we relax the constraints Eq.(\ref{bqdefinition}) the integrand still peaks for the same configurations that obey Eq.(\ref{bqdefinition}) strictly. Hence in the thermodynamic limit, approximating the integrand suitably around the peaking configurations we may reliably estimate the partition function. This reliability can be self consistently established by computing the variance or correlation functions.

The leading contribution to the functional is expected to come from the extremum which maximizes the integrand $Z_{site}$. Furthermore, since we are choosing to describe fluid phase of the model we seek such spatial configurations which are discrete translational and rotational invariant in auxiliary fields. The $Z_{site}$ over a space-independent field configuration $\tilde{\eta}, \tilde{\phi}$ is given by :
	\begin{equation}
		 \left. Z_{site} \right|_{{\eta = \tilde{\eta}}, { \phi = \tilde{\phi}}} = \displaystyle \left( 1 + 2 {\nu}e^{\displaystyle \ -i \tilde{\eta}} \text{cos} \tilde{\phi}  + {\lambda} e^{\displaystyle \ -2i \tilde{\eta}} + 90 {\mu} e^{\displaystyle \ 4 i \tilde{\eta}}  \right)
	\end{equation}

It is evident that the maximum for $Z_{site}$ occurs at $\tilde{\eta} = \tilde{\phi} = 0$ since all fugacities are positive. $Z_{site}$ at the maximum is given by $Z_{o}$ :
	\begin{equation}
		Z_{o} = (1 + 2 \nu + \lambda + 90 \mu)
	\end{equation}
The extremization with respect to $\tilde{\phi}$ is trivially true, while that with respect to $\tilde{\eta}$ yields, 
	\begin{equation}
		2 \nu + 2 \lambda = 4 (90 \mu) \label{mfsumrule}
	\end{equation}
This is a consequence of the sum rule Eq.(\ref{sumrule}) within the zeroth order approximation. Various densities are calculated from Eq.(\ref{observables}) with the approximate partition function $Z_{o}$. Furthermore, using Eq.(\ref{mfsumrule}) :
	\begin{subequations}
	\label{densities_meanfield}
	\begin{eqnarray}
		\textnormal{DB} & = & \frac{4 \nu}{2 + 5 \nu + 3 \lambda} 
		\\ \textnormal{HB} & = & \frac{2 \lambda}{2 + 5 \nu + 3 \lambda} 
	\end{eqnarray}
	\end{subequations}
	\begin{equation}
		\rho  =  \frac{\nu + \lambda}{2 + 5 \nu + 3 \lambda} \label{waterdensity_mf}
	\end{equation}
Eliminating $\lambda$ in Eqs.(\ref{densities_meanfield},\ref{waterdensity_mf}) we obtain :
	\begin{equation}
		\textnormal{HB} = 2 \rho - \frac{\nu}{\nu + 1} (1 - 3 \rho)	\label{equationofnetwork}
	\end{equation}
This is the ``equation of network" which indeed is a different way of writing the sum rule Eq.(\ref{sumrule}) in terms of dangling-bond fugacity $\nu$. $Z_{o}$ is given by :
	\begin{equation}
		Z_{o}  =  \left(1 + \frac{5}{2} \nu + \frac{3}{2} \lambda \right) =  \frac{1 + \nu}{1 - 3 \rho} = \frac{1}{1 - (5 \rho - \textnormal{HB})} \label{mf_Zo}
	\end{equation}
Rewriting it as the last term in Eq.(\ref{mf_Zo}) shows that it is also inverse of the density of void sites (every water molecule in our model necessarily occupies five sites and for every hydrogen-bond one site is double-counted and hence is compensated in the expression).

From the sum rule it follows $0 \leq \textnormal{HB} \leq 2 \rho$. Consequently, $\rho$ here varies between $\frac{\nu}{5 \nu + 2}$ and $\frac{1}{3}$. The upper bound on $\rho$ is indeed the highest possible density in the model while the lower bound is a consequence of MMF approximation, meaning that this description is self consistent only for densities greater than $\frac{\nu}{5 \nu + 2}$. Without loosing any generality, we choose $\tilde{\lambda} = 1$ or $\lambda = e^{\beta}$ i.e, measuring all energies in the units of hydrogen-bond energy. Then we make the observation that if temperature ($\frac{1}{\beta}$) is always positive, from Eq.(\ref{waterdensity_mf}) we can show that $\rho$ is greater than $\frac{1}{5}$. Furthermore, as $\beta \rightarrow \infty$, from Eqs.(\ref{densities_meanfield},\ref{waterdensity_mf}) we see that $\rho \rightarrow \frac{1}{3}$, $\text{HB} \rightarrow \frac{2}{3}$ and $\text{DB} \rightarrow 0$, showing that every water molecule's arms are all hydrogen-bonded leaving no void sites. This saturation at $\rho = \frac{1}{3}$ is verified to be exactly true by explicit construction of such configurations. We find that the highest density is not that of a unique crystal configuration. Instead, there are infinitely many configurations corresponding to different spatial and orientational arrangements of water molecules.

The equation of state in terms of densities is given by : 
	\begin{equation}
		\mathcal P_{o} = \frac{1}{\beta} \ln(Z_{o})  =  \frac{\ln(1 - 5 \rho + \text{HB})}{\ln(1 - 5 \rho + \text{HB}) - \ln(\text{HB}) } \label{equationofstate}
	\end{equation}	
where $\mathcal P_{o}$ is the pressure and $\beta$ is written in terms of natural energy units of the model. 

By considering the case where H and L arms are neither energetically favored or suppressed i.e., $\tilde{\nu} = 0$ and $\nu = 1$, the entropy per site is given by,
	\begin{eqnarray}
		 \nonumber S &=& -{\beta}^{2} \frac{\partial}{\partial \beta} \left( \frac{1}{\beta} \ln(Z_{o}) \right)
		\\ &=& \ln\left(\frac{7}{2} + \frac{3}{2}e^{\beta} \right) - \beta \frac{3 e^{\beta}}{{7}+ {3}e^{\beta}}   
	\end{eqnarray}
In the limit $\beta \rightarrow \infty$, $\rho$ reaches the maximum value $\frac{1}{3}$ and the entropy at the highest density tends to a constant value $\ln(\frac{3}{2})$. The constant $\frac{3}{2}$ compares exactly with that of Pauling's estimate for tetrahedral ice i.e., $2^2 \frac{6}{16} \equiv 1.5$ \cite{PaulingJACS1935} and agrees well with the numerical estimate by Nagle i.e., $1.50685 \pm 0.00015$ \cite{NagleJMathPhy1966}. We note that these results are independent of dimension, hence in two dimensions it also compares with the exact result of Lieb on `square ice' \cite{LiebPR1967Vol162}.

\subsection{Fluctuations} \label{oneloop}

Next, we evaluate the functional integral of the partition function by considering small fluctuations around the mean field and approximating them by a Gaussian i.e., one-loop correction \cite{ChaikinLubenskyCUP}. Expanding $Z_{site}$ up to quadratic terms, we obtain :
	\begin{widetext}
	\begin{eqnarray}
		\nonumber Z_{site} &\simeq & 1 \ + \ 2 {\nu} \ + \ {\lambda} \ + \ 90 {\mu}
		+ 2 \nu \left( - i \eta - \frac{{\eta}^{2}}{2} - \frac{{\phi}^{2}}{2} \right)
		+ \lambda \left( - 2 i \eta - 2 {\eta}^{2} \right)
		\\ \nonumber && + 90 {\mu} \left[ 4 i \eta + \frac{2}{3} i \sum_{\displaystyle \alpha} \nabla_{\alpha} \eta - \frac{8}{3} \left(\sum_{\alpha} \eta \nabla_{\alpha} \eta \right) - 8 {\eta}^{2}   \right.
		\\  \nonumber && \qquad \qquad \left.  - \frac{1}{5} \left(\sum_{\displaystyle \alpha} \nabla_{\alpha} \eta \right)^2 - \frac{2}{15} \sum_{\displaystyle \alpha} (\nabla_{\alpha} \eta)^{2} +  \frac{1}{15} \left(\sum_{\displaystyle \alpha} \nabla_{\alpha} \phi \right)^2 - \frac{2}{5} \sum_{\displaystyle \alpha} (\nabla_{\alpha} \phi)^{2} \right]
		\\  \nonumber & \simeq & Z_{o} \ \text{exp} \left\lbrace - i \eta(x)\left(\nu^{'} + 2 \lambda^{'} - 4 \mu^{'} \right) - \frac{2 i \mu^{'}}{3} \sum_{\alpha} \nabla_{\alpha} \eta  \right.
		\\ \nonumber && \left. \qquad \qquad - \frac{1}{2} \left[ \left( {\nu^{'}} + 4 \lambda^{'} + 16 \mu^{'} \right) \eta^{2}(x) + \frac{2\mu^{'}}{5}  \left(\sum_{\displaystyle \alpha} \nabla_{\alpha} \eta \right)^2 + \frac{4 \mu^{'}}{15} \sum_{\displaystyle \alpha} (\nabla_{\alpha} \eta)^{2}  \right. \right.
		\\  \nonumber && \left. \left. \qquad \qquad \qquad + \frac{16 \mu^{'}}{3}  \left(\sum_{\alpha} \eta  \nabla_{\alpha} \eta \right) - \left( (\nu^{'} + 2 \lambda^{'} - 4 \mu^{'}) \eta(x)  - \frac{2 \mu^{'}}{3} \sum_{\displaystyle \alpha} \nabla_{\alpha} \eta \right)^2 \right. \right.
		\\  && \qquad \qquad \qquad  \left. \left. +  {\nu^{'}} \phi^{2}(x) + \frac{4 \mu^{'}}{5}  \sum_{\displaystyle \alpha} (\nabla_{\alpha} \phi)^{2} - \frac{2 \mu^{'}}{15}  \left(\sum_{\displaystyle \alpha} \nabla_{\alpha} \phi \right)^2 \right] \right\rbrace
	\end{eqnarray}
	\end{widetext}
where $\nu^{'} = 2 \nu/Z_{o} $, $\lambda^{'} = \lambda/Z_{o}$, $\mu^{'} = 90 \mu/Z_{o}$ are the reduced fugacities; such that all of them are less than 1.

Inserting the above expression for $Z_{site}$ in Eq.(\ref{Zequation}) and evaluating the resulting Gaussian integral by Fourier transform in a periodic box the pressure $\mathcal P$ is given by :
	\begin{equation}
		\mathcal P =  \frac{1}{\beta} \left \lbrace \ln(Z_{o}) - \frac{1}{2} \int\limits^{\pi}_{-\pi} \frac{\displaystyle d^{3}k}{(2 \pi)^{3}} \left[ \ln(P_{\eta \eta}(\Delta)) + \ln(P_{\phi \phi}(\Delta))	\right]  \right \rbrace \label{pressure_oneloop}
	\end{equation}
where 
	\begin{eqnarray}
		P_{\eta \eta}(\Delta) &=& \left[ 64 \mu^{'}  \Delta^{2} \left( \frac{9}{10} - \mu^{'} \right) + 64 \mu^{'}\Delta  \left( - \frac{9}{10} + \mu^{'} - \frac{\nu^{'}}{4} - \frac{\lambda^{'}}{2} \right)  \right.
		\\ \nonumber &&   \qquad \qquad \displaystyle \left. + { \nu^{'}} + 4 \lambda^{'} + 16 \mu^{'} -  \left( \nu^{'} + 2 \lambda^{'} - 4 \mu^{'}  \right)^2  \right]
		\\ P_{\phi \phi}(\Delta) &=& \left[ \frac{\displaystyle 96 \mu^{'}}{\displaystyle 5}   \Delta \left( 1 - \Delta \right) + {\displaystyle \nu^{'}} \right]
	\end{eqnarray}	
and $\Delta = \displaystyle \frac{1}{6} \sum_{i=1}^{3} (1 - \text{cos}(k_{i}))$. Thereupon the densities $\rho$, HB, DB are calculated using Eq.(\ref{observables}). We compare these results with that of MC simulations in a later section (see sec.\ref{MMF_MC} on MC).

\subsection{Correlation functions} \label{correlations}

The position space correlation functions for $\eta(x)$ and $\phi(x)$ fluctuations i.e., $G_{\eta \eta}(r)$ and $G_{\phi \phi}(r)$, to the leading order, are given by :
	\begin{eqnarray}
		G_{\eta \eta}(r) \equiv \langle \eta(0) \eta(r) \rangle = \int\limits_{-\pi}^{\pi} \frac{d^{3}k}{(2 \pi)^3} \frac{e^{i \vec{k}\cdot\vec{r}}}{P_{\eta \eta}(\vec{k})} \label{Geta}
		\\ G_{\phi \phi}(r) \equiv \langle \phi(0) \phi(r) \rangle = \int\limits_{-\pi}^{\pi} \frac{d^{3}k}{(2 \pi)^3} \frac{e^{i \vec{k}\cdot\vec{r}}}{P_{\phi \phi}(\vec{k})} \label{Gphi}
	\end{eqnarray}

We note that to zeroth order $\mu^{'} \simeq \rho$, $\lambda^{'} \simeq \text{HB}$, $\nu^{'} \simeq \text{DB}$ and hence, using Eq.(\ref{mfsumrule}) :
	\begin{eqnarray}
		P_{\eta \eta}(\vec{k}) & \simeq & \displaystyle 64 \rho \left(\frac{9}{10} - \rho \right) 
		\left[ \left(\Delta -  \frac{9}{20 (\frac{9}{10} - \rho)} \right)^{2} + \frac{3 (\frac{9}{25} - \rho)}{8 (\frac{9}{10} - \rho)^{2}} \right] \label{Peta}
		\\ P_{\phi \phi}(\vec{k}) & \simeq & \displaystyle \frac{96 \rho}{5} \left[ \Delta (1 -\Delta)  + \frac{5 \text{(DB)}}{96 \rho} \right] \label{Pphi}
	\end{eqnarray}

The asymptotic behavior for large $r$ of $G_{\eta \eta}$ and $G_{\phi \phi}$ correlators can be obtained by pursuing small-$k$ expansion of the integrand and noting that for small $\vec{k}$, $\Delta \simeq \frac{1}{12} \sum_{i} k_{i}^{2} $. The $G_{\eta \eta}$ correlator for large $r$ is,
	\begin{equation}
		G_{\eta \eta}(r) \propto \frac{e^{-m_{1} r}}{r} \sin(\omega_{1} r)
	\end{equation}
where 
	\begin{eqnarray}
		m_{1} & = & \displaystyle \sqrt[4]{\frac{ \frac{3}{8}}{\frac{9}{10} - \rho } } \ \sin\left(\frac{1}{2}{\tan}^{-1} \sqrt{\frac{50}{27}\left(\frac{9}{25} - \rho \right) } \ \right)
		\\ \omega_{1} & = & \displaystyle \sqrt[4]{ \frac{ \frac{3}{8}}{\frac{9}{10} - \rho } } \ \cos\left(\frac{1}{2}{\tan}^{-1} \sqrt{\frac{50}{27}\left(\frac{9}{25} - \rho \right) } \ \right)
	\end{eqnarray}	
This shows $G_{\eta \eta}$ has periodic peaks reminiscent of the coordination shell peaks whose amplitudes are exponentially falling off.

The $G_{\phi \phi}$ correlator takes the following asymptotic form for large $r$, in addition to oscillatory behavior prominent at short distances :
	\begin{equation}
		G_{\phi \phi}(r) \propto \frac{e^{-m_{2} r}}{r} ; \qquad m_{2} = \sqrt{6 \left( \sqrt{1+ \frac{5 \ \text{DB}}{24 \rho}} - 1 \right)} \label{phicorrelation}
	\end{equation}
We make the observation from Eqs.(\ref{Peta},\ref{Pphi}) that the correlation functions Eqs.(\ref{Geta},\ref{Gphi}) of the mean field theory diverge if $\rho \rightarrow 0$ i.e., even the local fluctuations about the mean field are very large rendering the approximation invalid. Indeed the theory fails well before that because it violates the sum-rule already at $\rho = \frac{\nu}{5 \nu +2}$. This is the reason why our mean field theory fails for the gas phase and thus does not describe any liquid-gas transition. To sum it, in our model MMF approximation shows density saturation and the description is more accurate at higher densities only.  

Next, we compute some important physical correlations such as water-water and charge-charge.
	\begin{equation}
		\langle W(x) W(y) \rangle = \left\langle \frac{\mu \lbrace \ldots \rbrace}{Z_{site}(x)} \frac{\mu \lbrace \ldots \rbrace}{Z_{site}(y)} \right\rangle
	\end{equation}
The non-zero value of $W$-field at each site $x,y$ picks only the term proportional to the fugacity $\mu$ in $Z_{site}$ expression Eq.(\ref{Zsiteequation}) at both sites. The $\lbrace \ldots \rbrace$ include same set of terms as in Eq.(\ref{Zsiteequation}). We now make the one-loop approximation of retaining terms only upto quadratic in fields and further, subtract out single-site averages i.e., $\langle W(x) W(y) \rangle - \langle W(x) \rangle  \langle W(y) \rangle$. Thereby, correlation of water fluctuations to leading order is given by :
	\begin{equation}
		\langle W(0) W(r) \rangle - {\rho}^{2}  \simeq  - \rho^{2} \displaystyle \int\limits_{-\pi}^{\pi} \frac{d^3 k}{(2 \pi)^3} \ \frac{(4 - 8\Delta(1-\rho))^{2} e^{i \vec{k} \cdot \vec{r}}}{P_{\eta \eta}(\vec{k})} \label{wwcorrelation}
	\end{equation}
A similar computation for the $q(x)$ field can be done, wherein the non-zero $q$ value picks terms proportional to the fugacity $\nu$ in $Z_{site}$ at either sites (Eq.\ref{Zsiteequation}). To the leading order,
	\begin{equation}
		\langle q(0) q(r) \rangle  \simeq  - (\text{DB})^{2} \displaystyle \int\limits_{-\pi}^{\pi} \frac{d^3 k}{(2 \pi)^3} \  \frac{e^{i \vec{k} \cdot \vec{r}}}{P_{\phi \phi}(\vec{k})} \label{qqcorrelation}
	\end{equation}
Note that the single-site average over $q(x)$ field is zero. We remark that in this model $W(x)$ being charge neutral gets contribution from the neutral $\eta(x)$ field, while $q(x)$ from that of the charged $\phi(x)$ field.

\subsection{Coulomb interaction} \label{coulomb}

In this section we consider the influence of long range Coulomb interaction potential between the charges H, L given in terms of charge $e$ as :

	\begin{equation}
		H_{Col} = \frac{e^2}{2} \displaystyle \sum_{x,y}^{'} \sum_{\alpha, \gamma} \frac{ H_{\alpha}(x) H_{\gamma}(y) }{|x + e_{\alpha} - y - e_{\gamma}|}
	\end{equation}
where prime over summation means $x \neq y$. In our model we envisage the charges at the tip of water's H or L arms. This can be incorporated in our analysis by using an auxiliary field technique :
	\begin{eqnarray}
		e^{- \beta H_{Col}}  & \rightarrow & \displaystyle \sqrt{{\text{det}(-\square + m^2)}} \prod_{x} \int \frac{d\chi(x)}{\sqrt{2 \pi}}
		\\ \nonumber && \qquad \text{exp} \sum_{x} \left[{-\frac{1}{2} \chi(x)(-\square + m^2)\chi(x) + i  \sqrt{\beta e^{2}} \sum_{\alpha} H_{\alpha}(x + e_{\alpha}) \chi(x)} \right]
	\end{eqnarray}
where the laplacian operator $\square \chi(x) = \sum_{\alpha} \left(\chi(x+e_{\alpha}) - \chi(x) \right) = \sum_{\alpha} \nabla_{\alpha} \chi(x)$ and $m$ is a parameter that regulates the range of interaction. The interaction potential behaves as $\frac{e^{-mr}}{r}$ for large distances $r$, which also reduces to Coulomb interaction when $m = 0$. If the lattice constant and $m$ are both taken to be zero, then it reduces to exact Coulomb interaction for all $r$.
	
By inserting the above in our partition function Eq.(\ref{Zequation}) all the interactions of water degrees of freedom remain unchanged with the following transformation $\eta \rightarrow \eta$, $\phi \rightarrow \phi +  \chi \sqrt{\beta e^{2}} $. The extremum of the new partition function is still at $\tilde{\eta} = \tilde{\phi} = \tilde{\chi} = 0$. The leading zeroth order term remains unchanged; the one-loop correction about the mean field gets additional contributions due to quadratic terms corresponding to $\phi \chi$ and $\chi \chi$ in the Gaussian expansion, given by : 
	\begin{eqnarray}
		P_{\chi \chi} & = & \left( {12} \Delta + {m^2} \right) + \frac{96 \mu^{'} {\beta e^2}}{5}  \Delta \left( 1 - \Delta \right)
		\\
		P_{\phi \chi} & = & \frac{96 \mu^{'} \sqrt{\beta e^2}}{5} \Delta \left( 1 - \Delta \right)
	\end{eqnarray}

The pressure $\mathcal P$ with Coulomb interactions is,
	\begin{eqnarray}
		\mathcal P  =   \frac{1}{\beta} \left \lbrace \ln(Z_{o}) + \frac{1}{2}\ln(12 \Delta + m^{2}) - \frac{1}{2} \int\limits^{\pi}_{-\pi} \frac{\displaystyle d^{3}k}{(2 \pi)^{3}} \left[ \ln(P_{\eta \eta}(\Delta)) + \ln(P(\Delta)) \right] \right \rbrace 
	\end{eqnarray}
where
	\begin{eqnarray}
		P(\Delta) & \equiv & P_{\phi \phi} P_{\chi \chi} - P_{\phi \chi}^{2}
		\\  \nonumber & = & \frac{96 \mu'}{5} (12 \Delta + m^2) \left[ \Delta \left( 1 - \Delta \right) + \frac{5 \nu'}{96 \mu'} + \frac{ \beta e^2 \nu' }{12 \Delta + m^2} \Delta \left( 1 - \Delta \right)  \right]
	\end{eqnarray}	
which to the leading order can be written as :
	\begin{equation}
 		P(\vec{k}) \simeq  \displaystyle \frac{96 \rho}{5} (12 \Delta + m^2) \left[\Delta (1 - \Delta) + \frac{5 \text{(DB)}}{96 \rho} + \left(\frac{\beta e^2 \text{(DB)} }{12 \Delta + {m^2}} \right) \Delta (1 - \Delta) \right]
	\end{equation}

The two-point correlations of the fields in momentum space are given by :
	\begin{subequations}
	\begin{eqnarray}
		\langle \phi(\vec{k}_1) \phi(\vec{k}_2) \rangle & = & \frac{P_{\chi \chi}(\vec{k}_1) \delta({\vec{k}_1+\vec{k}_2})}{P(\vec{k}_1)}
		\\ \langle \chi(\vec{k}_1) \chi(\vec{k}_2) \rangle & = & \frac{P_{\phi \phi}(\vec{k}_1) \delta({\vec{k}_1+\vec{k}_2})}{P(\vec{k}_1)}
		\\ \langle \phi(\vec{k}_1) \chi(\vec{k}_2) \rangle & = & \frac{P_{\phi \chi}(\vec{k}_1) \delta({\vec{k}_1+\vec{k}_2})}{P(\vec{k}_1)}
	\end{eqnarray}
	\end{subequations}	
	
The factor $e^{2}$ being small the strength of Coulomb interactions in comparison with other interactions is weak. Furthermore, all charge effects are proportional to $\beta e^{2} \text{(DB)}$ wherein $\beta \text{(DB)}$ is always finite since as $\beta \rightarrow \infty$, $\text{DB} \rightarrow 0$. Consequently, effect of Coulomb interactions on thermodynamic properties such as equation of network and equation of state is small. Correlation functions are not perturbed very much at large distances. It does, however, shift the coordination peaks slightly.

\section{Monte Carlo simulation} \label{MC}

The hydrogen-bond network model with all the accompaning constraints is simulated in the background of a cubic lattice using standard Monte Carlo (MC) procedure for grand canonical ensemble i.e. $\tilde{\mu} V T$ ensemble \cite{BinderLandauCUP, FrenkelSmitAP}. The moves implemented at each simulation step on a randomly chosen site are :
	\begin{itemize}
		\item if there is no water, ``insert" one provided there are four arms free around the site
		\item if there is water, with equal probability either ``delete" water or ``rotate" the arms to another possible allowed configuration as dictated by Eq.(\ref{Hamiltonianredefinition}).
	\end{itemize}
The new configuration is accepted with a probability of Boltzmann weight over energy change in compliance with detailed balance condition \cite{FrenkelSmitAP,WhitehouseJPhyA1984}. All energies $\beta, \tilde{\nu}, \tilde{\mu}$ are calculated in units of hydrogen-bond energy i.e., $\tilde{\lambda} = 1$. Furthermore, we considered specifically $\tilde{\nu} = 0$ case thereby making the theory essentially parameter-free other than temperature and chemical potential which are varied in steps as per needs of the simulation.

The thermodynamic observables of the $\tilde{\mu} V T$ ensemble i.e., number of water molecules $\rho$, number of hydrogen bonds $\text{HB}$, total energy $E$ are obtained as averages from the simulation. A significant equilibriation time is allowed before the production run begins. In the production run the configurations are sampled about every $50-300$ MC steps over a simulation time of $10^{5}-10^{6}$ MC steps in order to carry out averaging procedure. Running averages computed at every sampling step allow determination of the efficiency of sampling procedure over the extent of simulation time. It is ensured that there is no observable overall rise or fall in the averages and variances and that the results smoothly converge to within a relative error of $10^{-2}-10^{-3}$. Then the same are noted down for record. At every step various configuration' statistics are recorded such as number of molecules with one, two, three, and four hydrogen bonds. The on-the-run variances in all the quantities are computed using standard discrete sum formula.

The size dependence of the averages is ascertained and an optimal lattice size of $20$ sites per side is found to be closely reproducing averages upto fourth decimal relative to bigger lattice sizes.

A `successive seeding' procedure is employed to adiabatically progress in the parameter space of the model. For example, to carry out simulations at a fixed temperature and a range of chemical potentials an initial run carried out at low densities and equilibriated over long time is used as a `seed' i.e., input configuration for the successive run. This is repeated to reach higher densities. This procedure accelerates equilibration considerably compared to any random seed configuration. Furthermore, since a range of chemical potentials are explored the step size in the parameter value is appropriately adjusted such that a quench-like situation is avoided. Thereby, the system evolves smoothly in configuration space without any unwanted domains persisting. The end averages remain unchanged when the seeding procedure is carried out in an alternative parameter space, for example, instead of chemical potential temperature can be varied in small steps. This confirms the absence of any possible bias created by our successive seeding procedure in most part of the parameter space (except near first order phase transitions where hysteresis exists).

\begin{figure*}
		\epsfig{file=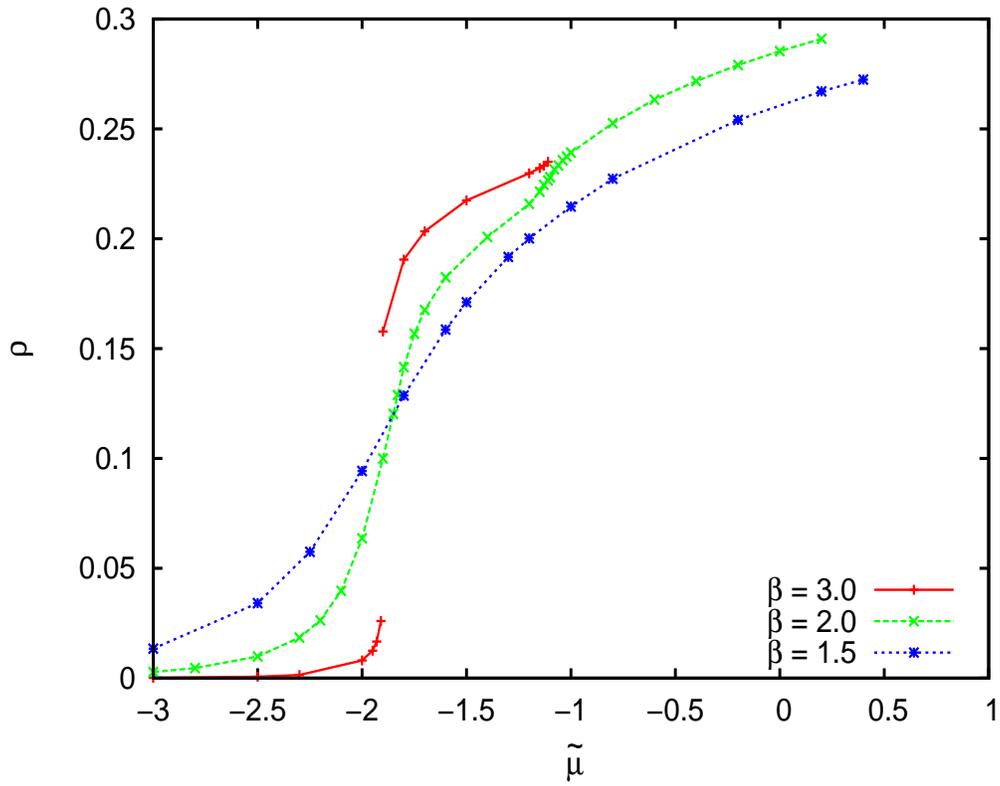,width=0.85\linewidth,height=0.45\textheight}
		\caption{\label{fig_rhomu} First order phase transition in MC simulation seen for $\beta > 2$: isotherms are $\beta = 3.0, 2.0, 1.5$; (red) $\beta = 3.0$ isotherm discontinuity at $\tilde{\mu} = -1.91$, density changing from $\rho \sim 0.025$ ($h \sim 1.2$) to $\rho \sim 0.16$ ($h \sim 1.7$).}
	\end{figure*}

In order to obtain equation of state using MC we employed fixed temperature simulations, followed by application of Gibbs-Duhem procedure to obtain pressure versus density curves \cite{PanagioJPhy2000Vol12, PabloARPC1999, MauricioJCP2007Vol126}. In the $\tilde{\mu} V T$ ensemble temperature is fixed and a range of chemical potentials are explored starting from zero density where pressure can be normalized to the value zero. Employing the successive seeding procedure we obtain various thermodynamic averages as functions of chemical potential. The Gibbs-Duhem procedure involves integrating the $\rho$ \text{vs.} $\tilde{\mu}$ curve so that pressure can be obtained from the following relation :
	\begin{equation}
		\mathcal P = \int\limits_{\tilde{\mu}_{i}}^{\tilde{\mu}_{f}} \rho(\tilde{\mu}) \ d\tilde{\mu}
	\end{equation}
where the $\tilde{\mu}_{i}$ corresponds to $\rho = 0$ and $\mathcal P = 0$ and $\tilde{\mu}_{f}$ to that of the desired $\rho$. A set of representative $\rho$ \textnormal{vs.} $\tilde{\mu}$ curves are shown in Fig.(\ref{fig_rhomu}). Each such curve has two prominent shoulders where considerable slope change occurs. At very high $\tilde{\mu}$ values no more equilibriated configurations could be traced and the density starts shooting up to the saturation value.

We also made preliminary investigation in studying phase transitions in the model. As shown in  Fig.(\ref{fig_rhomu}) the model exhibits discontinuity in $\rho(\tilde{\mu})$ for temperatures $\beta > 2$ (in units of hydrogen-bond energy). For instance, at $\beta = 3$ the system jumps from a low dense ($\rho \sim 0.025, h \sim 1.2$) to higher density ($\rho \sim 0.16, h \sim 1.7$) where $h = \frac{\text{HB}}{\rho}$ is average number of hydrogen-bonds per molecule. Hysteresis is also observed when the system progresses first from low-dense to high-dense state upon increasing $\tilde{\mu}$ and then retracing the path by decreasing $\tilde{\mu}$. This indicates the presence of first-order phase transition in the region. We interpret this as liquid-gas transition in the model.

\subsection{MMF and MC} \label{MMF_MC}

One of the important expositions of the MMF theory is the ``equation of network", a functional relationship between total water density and HB density. From Eq.(\ref{equationofnetwork}), at $\nu = 1$, the mean field equation of network is given by,
	\begin{equation}
		\text{HB} = \frac{7 \rho}{2} - \frac{1}{2} \qquad \text{for } \rho \geq \frac{1}{5} \label{networkequation}
	\end{equation}
In MC simulation we identify the liquid-gas transition for temperatures $\beta > 2$. As shown in Fig.(\ref{fig_rhomu}), for $\beta = 3.0$ at $\tilde{\mu} \simeq -1.91$ water density jumps from $0.025$ to $0.16$. The jump gap increases with $\beta$, so does the higher density state to which system jumps. Within our limited exploration of the phase diagram we identify that water densities $\rho \gtrsim 0.16$ correspond to liquid phase, only above which MMF theory is self-consistent (Eq.\ref{networkequation}). Furthermore, the linear relationship between $\rho$ and HB is accurate at high pressures as confirmed by MC simulation. As noted earlier, MMF approximation complies with the sum rule of the system; the same is exactly satisfied by MC simulation at every configurational move and hence over ensemble average. Thus, the agreement between mean field (zeroth order and one loop correction) and MC over the equation of network is inescapable. However, since the mean field description holds forte in thermodynamic limit, the HB density in the above equation corresponds to lowest free energy (or high pressure) states for each water density.

\begin{figure*}
		\epsfig{file=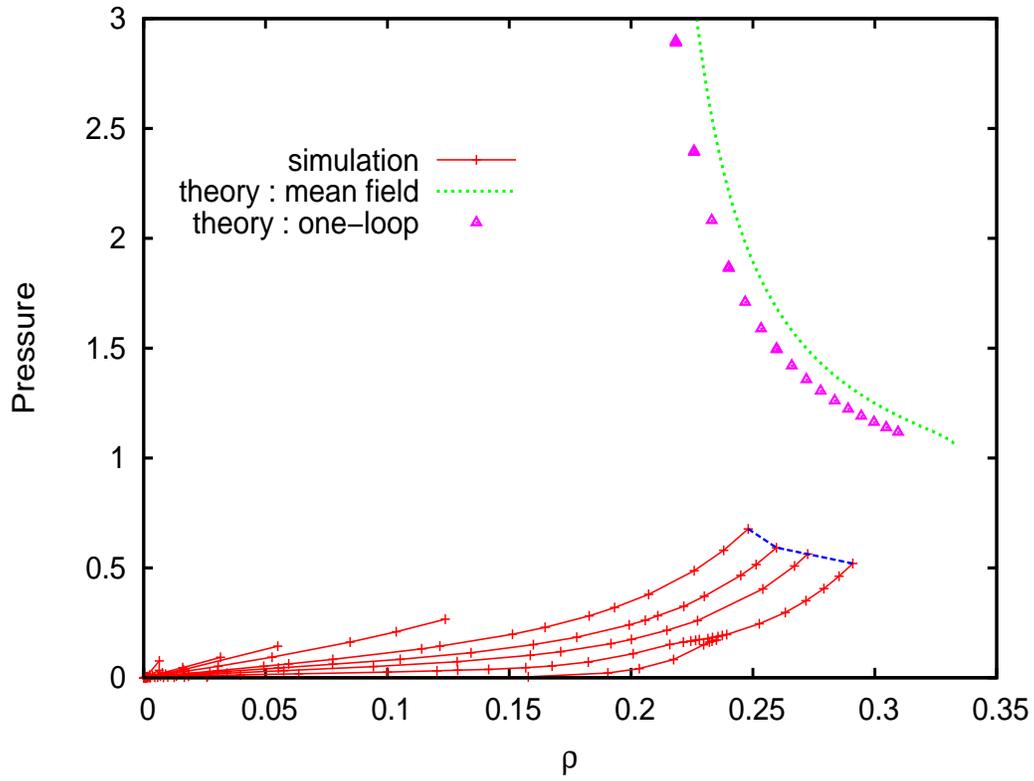,width=0.85\linewidth,height=0.45\textheight}
		\caption{\label{fig_equationofstate} Equation of state : (green, dotted line) MMF zeroth order; (magenta, filled triangles) with one-loop correction; (red lines) isotherms (temperature increasing from top to bottom) in MC simulation; (blue, dashed line) locus of high pressure states.}
	\end{figure*}

Next, we compare the MMF computation of equation of state with that of MC simulation. From theory, at $\nu = 1$, 
	\begin{equation}
		\mathcal P_{o} = \frac{\ln\left(\frac{2}{1 - 3 \rho}\right)}{\ln\left(\frac{7 \rho - 1}{1 - 3 \rho}\right)} \qquad \text{for } \rho \geq \frac{1}{5}
	\end{equation}
The comparison is put forth in Fig.(\ref{fig_equationofstate}). It shows that the locus of high pressure states at each density in MC simulation has the same profile as in MMF theory (both in zeroth order and one-loop'). Further, the qualitative agreement is good only in the high density regime where the equation of network too is accurate. A quantitative comparison of equation of state between MMF and MC is unreliable because MMF pressure absurdly vanishes at $\rho = \frac{1}{7}$ whereas, physically, the pressure is zero in this model only at $\rho = 0$. As discussed earlier, MMF approximation fails for small densities. Therefore, a consistent normalization between various schemes of calculation is not present. Thus, the qualitative picture obtained from MMF calculation is only indicative, nevertheless consistent with MC results.

The spatial correlation functions are also computed from the MC simulation and compared with that of analytical expressions obtained within the mean field approximation. Firstly, we note that the underlying hyper-cubic lattice dominates all correlations especially at short distances. Then, to extract the rotational invariant part of any function $f(x)$ we implement the following projection procedure :
	\begin{subequations}
	\label{projectionequation}
	\begin{eqnarray}
		R(r) & = & \displaystyle \sum_{\vec{x}} {\Theta}\left(|\vec{x}| - r \right) \  {\Theta}\left( \left(r + \delta \right) - |\vec{x}| \right)
		\\ f(r) & = & \displaystyle \frac{1}{{R(r)}} {\displaystyle \sum_{\vec{x}} f(\vec{x}) \ \Theta\left(|\vec{x}| - r \right) \ \Theta\left( \left(r + \delta \right) - |\vec{x}| \right)}
	\end{eqnarray}
	\end{subequations}
where $|\vec{x}| = \sqrt{\sum_{i} x_{i}^{2}}$ and $\Theta$ is Heaviside step function defined as $\Theta(x-a) = 1$ for $x \geq a$ and $0$ for $x < a$. The important correlations in the model are $<W(0) W(r)>$ and $<q(0) q(r)>$.  The $k$-space integrals in Eqs.(\ref{wwcorrelation}, \ref{qqcorrelation}) are computed numerically and the MMF correlation functions are compared with those of MC simulation. The positions of coordination peaks in both are in agreement. In addition, charge correlations show similar exponential fall-offs asymptotically. Fluctuations are seen to increase as we approach saturation density in both cases, however they are higher in simulation.

\section{MMF, Experiments and MD} \label{MMF_expt}

MMF theory predicts correlation length of charge fluctuations i.e., in $\langle q q \rangle$ correlation function Eqs.(\ref{phicorrelation},\ref{qqcorrelation}). These fluctuations could give rise to long distance correlations in other charged fields in the system such as in dipole, as seen in MD simulations (wherein two correlation lengths of order $5.2$ \AA{} and $24$ \AA{} have been observed, former being $10$ times larger than the latter in strength) \cite{KanthPRE2010Vol81}. To relate to MMF we make the following observations. Water molecule on average participates in $3.6$ hydrogen-bonds in ambient conditions \cite{JorgensenMolPhy1985Vol56, MartiJCP1996Vol105, MartiPRE2000Vol61} i.e., $\frac{\text{DB}}{\rho} = 0.4$. Using this value in $m_
{2}$ of Eq.(\ref{phicorrelation}) the correlation length turns out to be approximately $2.02 \times $ lattice constant in MMF. We next estimate the lattice constant in physical units by comparing the position of highest coordination peak in $\langle W(0) W(r) \rangle$ correlation obtained from MMF to that of experimental data. This procedure gives a lattice constant equal to $\frac{2.8}{1.75}$ \AA{}. Hence in physical units MMF predicts a correlation length of about $3.2$ \AA{} for liquid water. It should be noted that these predictions are not robust as the coefficients such as $\frac{5}{24}$ in Eq.(\ref{phicorrelation}) might vary with topology of the underlying lattice.

\begin{figure*}
		\epsfig{file=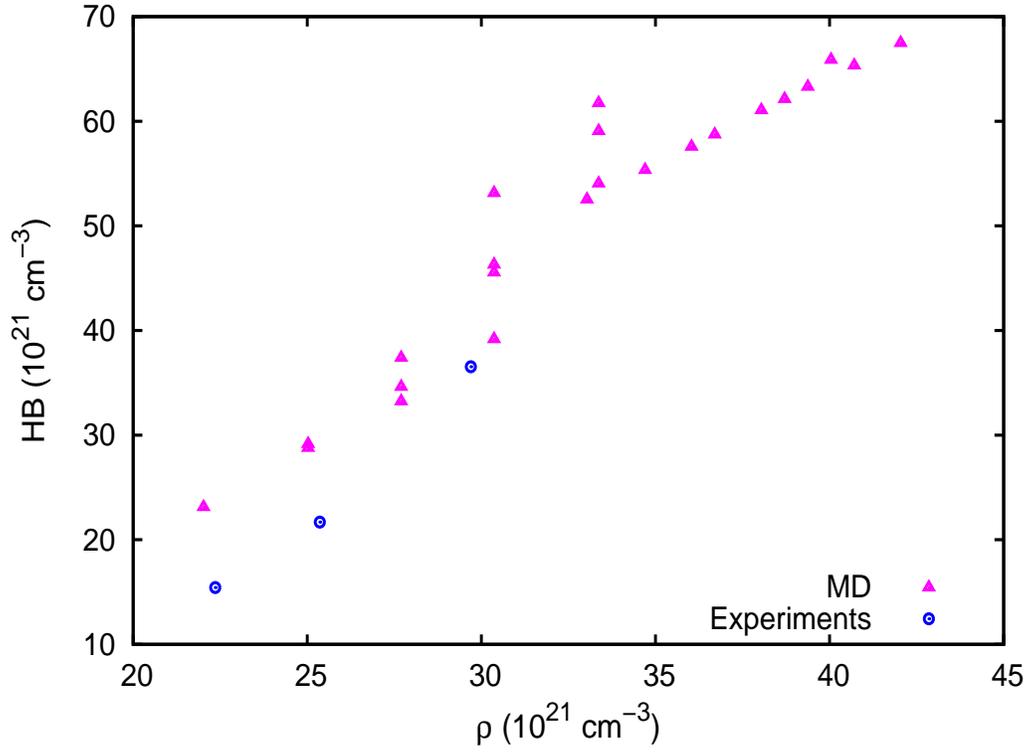,width=0.85\linewidth,height=0.45\textheight}
		\caption{\label{fig_equationofnetwork_expt} Equation of network : (blue, filled circles) Experiments and (magenta, filled triangles) MD simulations.}
	\end{figure*}

We also pursued a preliminary comparison of the equation of network from experiments and MD simulations. The density of hydrogen bonds is indirectly probed and inferred in various experiments \cite{TokushimaCPL2008Vol460,SkinnerCPL2009Vol470,BrubachJCP2005Vol122,PatrickPRB2007Vol76,KingJCP1977Vol67} and also computed in MD and MC simulations under varying external conditions \cite{MartiPRE2000Vol61,MartiJCP1996Vol105,KalinichevJMolLiq1999Vol82,BlumbergJCP1984Vol80}. The latter works used different definitions for hydrogen-bond computation such as energy-based, geometry-based or hybrid. The data is put in the perspective by converting the mass densities to number densities using known radius of a water molecule. As shown in Fig.(\ref{fig_equationofnetwork_expt}), in the region of high molecular density i.e., corresponding to liquid water we find that HB density and water density are linearly related to each other. A linear fit function is used for HB vs. $\rho$ curve and compared with that of the MMF equation Eq.(\ref{equationofnetwork}). We find that the dangling bond fugacity $\nu$ is in the range $(0.06,0.18)$, implying that the corresponding energy $\tilde{\nu}$ is positive and large compared to thermal energy i.e., dangling bonds are highly disfavored in liquid water. 
 
Another useful quantity namely fraction of water molecules with $i$ hydrogen bonds can also be calculated. Consider a water molecule in a configuration where its $i$-number of arms are hydrogen-bonded to neighboring molecules and its other $(4-i)$ arms remain dangling type. A weight can be associated with each such configuration defined in terms of appropriate site fields and summed over all possible orientations of the molecule. We denote this weight averaged with respect to the full partition function for each $i$ as $p_{i}$.  For instance, the averaged weight assigned to a molecule which is hydrogen-bonded to only two other molecules is given by :
	\begin{eqnarray}
		\nonumber p_{2} &=& \sum^{'}_{\lbrace \alpha_{1}, \alpha_{2}, \ldots, \alpha_{6} \rbrace} \left\langle W(x) \ \delta({b(x + e_{\alpha_{1}}),0}) \ \delta({b(x + e_{\alpha_{2}}),0}) \ \delta({b(x + e_{\alpha_{3}}),2}) \ \delta({q(x + e_{\alpha_{3}}),0}) \right.
		\\ && \left. \qquad \delta({b(x + e_{\alpha_{4}}),2}) \ \delta({q(x + e_{\alpha_{4}}),0}) \ \delta({b(x + e_{\alpha_{5}}),1}) \ \delta({b(x + e_{\alpha_{6}}),1}) \right\rangle
	\end{eqnarray}
The prime over summation means dissimilar $\alpha$. The probability for a $i$-bonded molecule at any site $x$ is probability that any two directions around central site have zero arms, each denoted by $\delta({b(x + e_{\alpha}),0})$, that $i$ other directions have a hydrogen-bond denoted by $\delta({b(x + e_{\alpha}),2}) \ \delta({q(x + e_{\alpha}),0})$ and that remaining $(4 - i)$ sites are of dangling bond denoted by $\delta({b(x + e_{\alpha}),1})$. The summation over the set ${\lbrace \alpha_{1}, \alpha_{2}, \ldots, \alpha_{6} \rbrace}$ implies summing over all possible rearrangements of hydrogen-bonds, dangling bonds among all the directions. With $\left[ 6 \atop 2 \right]$ ways of choosing empty site, $\left[ 4 \atop 2 \right]$ ways for two hydrogen bond sites the $p_{2}$, to the leading order, is given by :
	\begin{equation}
		p_{2} \simeq \left[ 6 \atop 4 \right]  \left[ 4 \atop 2 \right]  \text{(HB)}^{2} \text{(DB)}^{2}  = 15 (6) \rho^{4} {h}^{2} (2 - h)^{2}
	\end{equation} 
where $h = \frac{\text{HB}}{\rho}$ is average number of hydrogen bonds per molecule. Similarly, other $p_{i}$'s can be enumerated and computed upto leading order. For $i = 0, \ldots , 4$, 	
	\begin{equation}
		p_{i} \simeq 15 \left[ 4 \atop i\right] \rho^{4}  {h}^{i} (2 - h)^{4-i}
	\end{equation} 
Thereupon, $f_{i}$ which is fraction of $i$-bonded molecules can be calculated from the relation, 
 	\begin{eqnarray}
		  f_i  =  \displaystyle \frac{V p_{i}}{\sum_{i=0}^{4} V p_{i}}
		   \simeq  \frac{1}{2^4} \left[ 4 \atop i\right]  h^{i} (2-h)^{4-i}
	\end{eqnarray}
Note that the above expression is obtained to zeroth order approximation within the model. There exist one-loop corrections to it that can be calculated from the MMF theory, but numerically they are small. These distributions agree very well with MD simulations \cite{BlumbergJCP1984Vol80}. Further, probabilities for various cluster configurations (n-mers) can also be calculated within the MMF theory along the same lines as above. Liquid water is known to form molecular clusters such as trimers, tetramers, pentamers  \cite{SaykallyScience2004Vol306}. 

\section{Discussion} \label{discussion}

We analyzed the statistical mechanics of molecules with attractive bonding interactions. We argued that these systems typically have sum rules built-in which have to be respected in the analysis. In addition, there are other constraints, such as those shown in Fig.(\ref{disallowedconfigs}) for water, all of which are captured by defining fields such as $b(x)$ and $q(x)$ as in Eq.(\ref{bqdefinition}). The analysis yields ultimately a functional integral in terms of auxiliary fields $\eta(x)$, $\phi(x)$ which are conjugates of the discrete fields $b(x)$ and $q(x)$ respectively. A simple-minded mean field analysis of the auxiliary fields is shown to reproduce all major thermodynamic properties of the system, which are verified to be reasonably accurate in liquid phase and near saturation density. Principally, ``equation of network" which is another manifestation of the ``sum rule" plays an important role. We showed that all these consequences are known to be qualitatively true in experiments and MD simulations as well.

The MMF approximation is shown to be the leading order in a formal Feynman loop expansion of the functional integral with necessarily small parameters called reduced fugacities. Correlation functions and their properties have been computed and shown to qualitatively agree with MC simulation. It is however observed that many of the details in correlation functions do depend upon the underlying lattice, but the analytic structure is amenable to interpretation in the continuum as well. Pressure and correlation lengths have large contribution from the fluctuations of DB and $\rho$ in the form $\frac{\text{DB}}{\rho}$. We note that any long range interactions such as Coulombic have very little quantitative effect on our results.

The MMF theory is seen to be inconsistent at low densities, consequently fails to describe the gas phase of the model and the corresponding liquid-gas phase transition. In the explicit model considered here we do not have a crystalline solid phase, but we may have a glassy phase. Namely, at zero temperature the lowest energy configuration is infinitely degenerate. In this region MMF theory does not show any liquid-glass first order phase transition, however it does indicate a second order transition as seen from $q(x)$ correlator. In Eq.(\ref{phicorrelation}) for $\beta \rightarrow \infty$: $\rho \rightarrow \frac{1}{3}$, $\textnormal{DB} \rightarrow 0$ implying $m_{2} \rightarrow 0$. In MC simulation we did observe dynamical slowing down as we reach saturation density, perhaps indicative of a phase transition.

The MMF technique can be applied to associating liquids which may constitute variety of molecules and hence with variety of sum rules. Therefore, we expect a variety of dangling bond fluctuations contributing to pressure and correlation lengths. Indeed, there may even be variety of phase transitions. Essentially, the MMF technique is amenable to analysis of all such systems.

\section{Acknowledgments}
We acknowledge discussions with Prof. Gautam Menon on MC simulation data analysis.

\pagebreak
\section{Figures}

	\begin{figure*}
		\epsfig{file=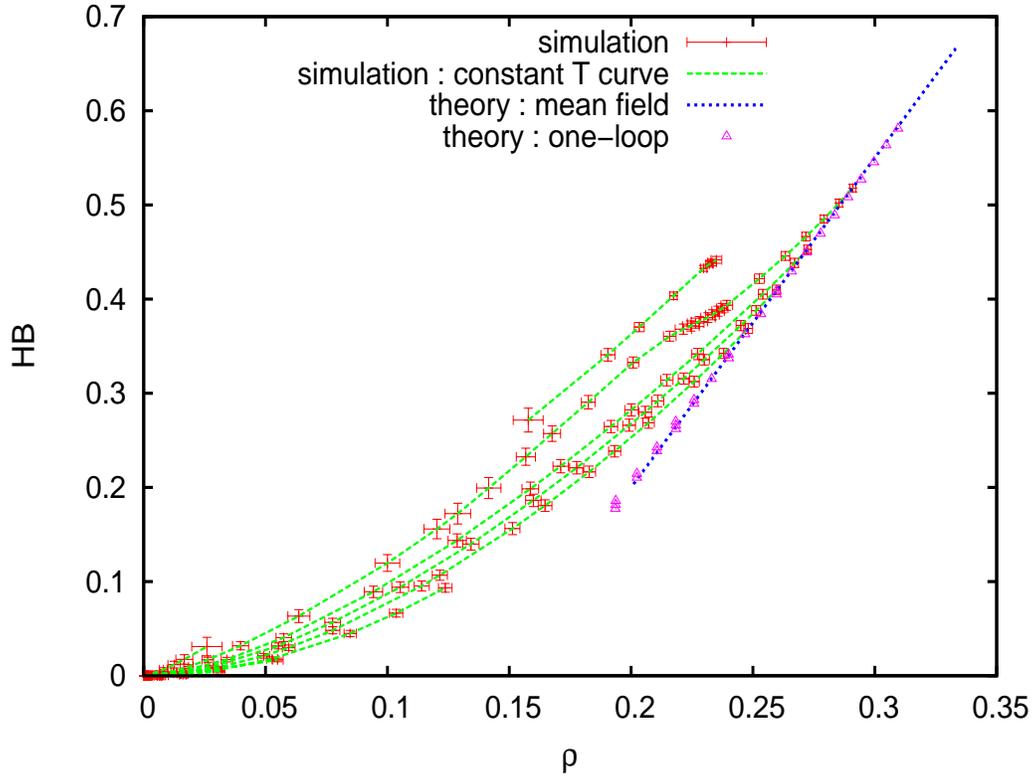,width=0.85\linewidth,height=0.45\textheight}
		\caption{\label{fig_equationofnetwork} Equation of network for $\tilde{\nu} = 0$ : (green, dashed lines) MC simulation isotherms (temperatures increasing from bottom to top); (blue, dotted line) mean field equation; (magenta, open triangles) with one-loop correction. }
	\end{figure*}

	\begin{figure*}
		\epsfig{file=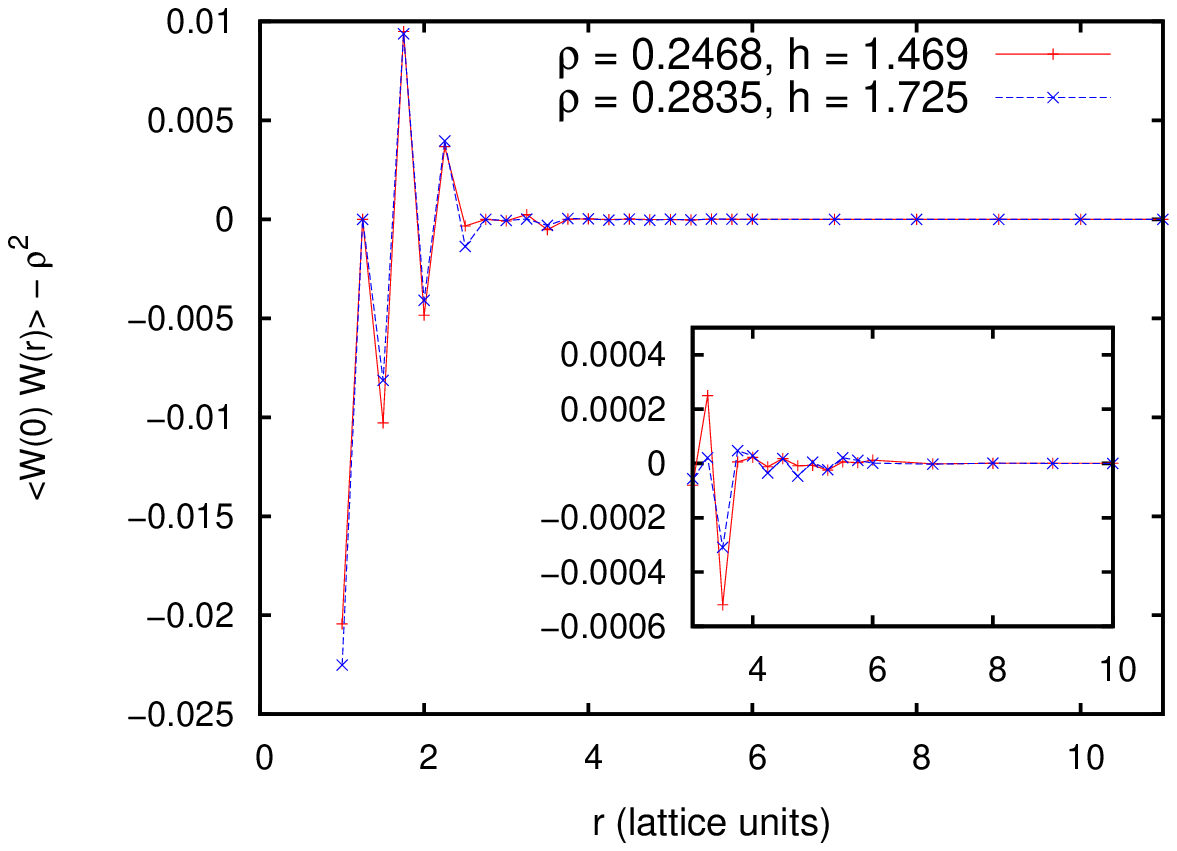,width=0.65\linewidth,height=0.4\textheight}
	    \caption{\label{fig_thww} MMF : water fluctuations' correlation function }
	\end{figure*}

	\begin{figure*}
		\epsfig{file=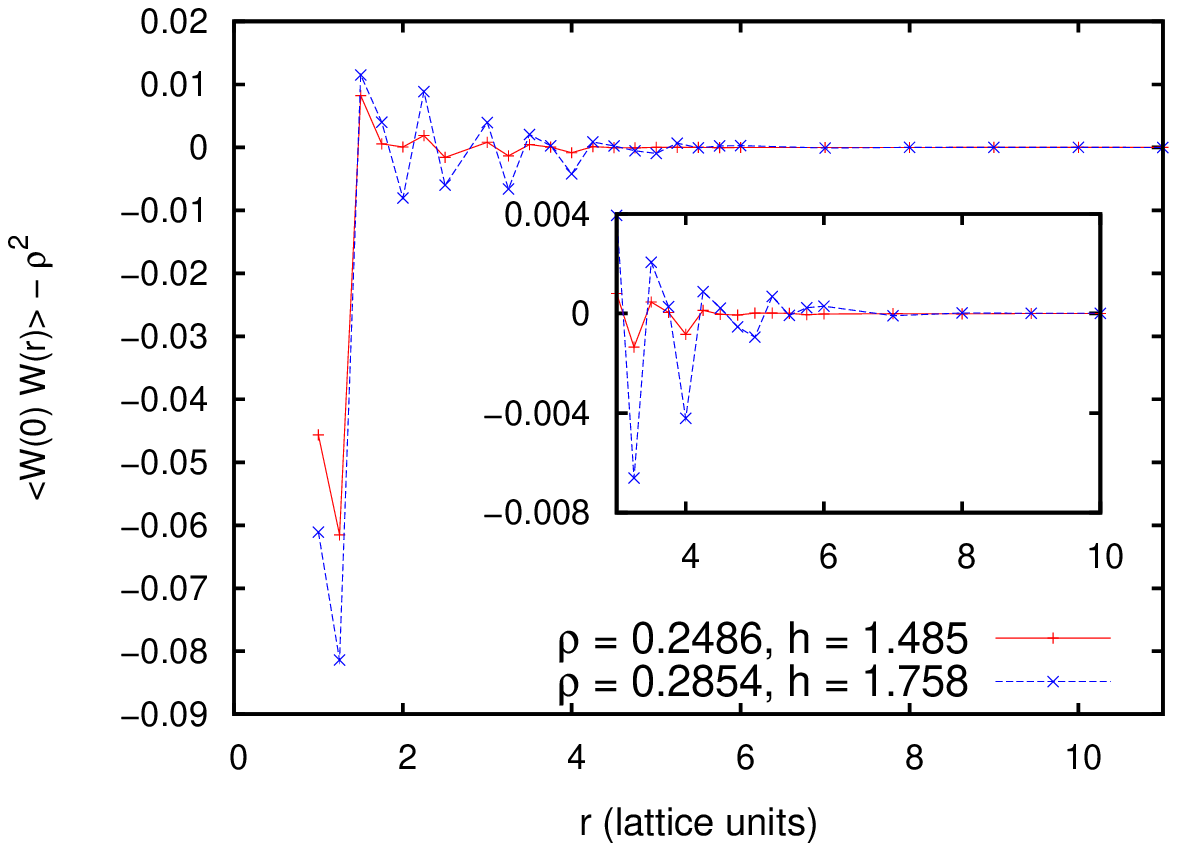,width=0.65\linewidth,height=0.4\textheight}
		\caption{\label{fig_simww} MC : water fluctuations' correlation function }
	\end{figure*}

	\begin{figure*}
		\epsfig{file=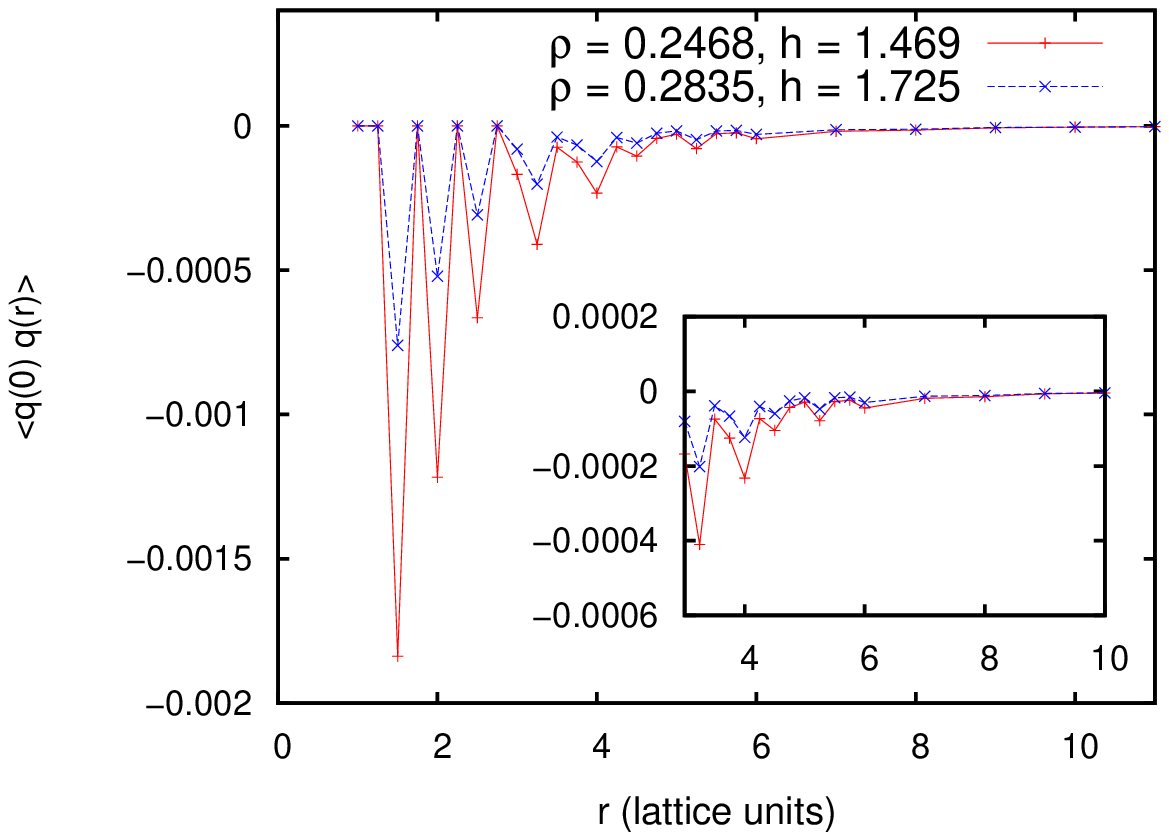,width=0.65\linewidth,height=0.4\textheight}
	    \caption{\label{fig_thqq} MMF : charge fluctuations' correlation function}
	\end{figure*}
	
	\begin{figure*}
		\epsfig{file=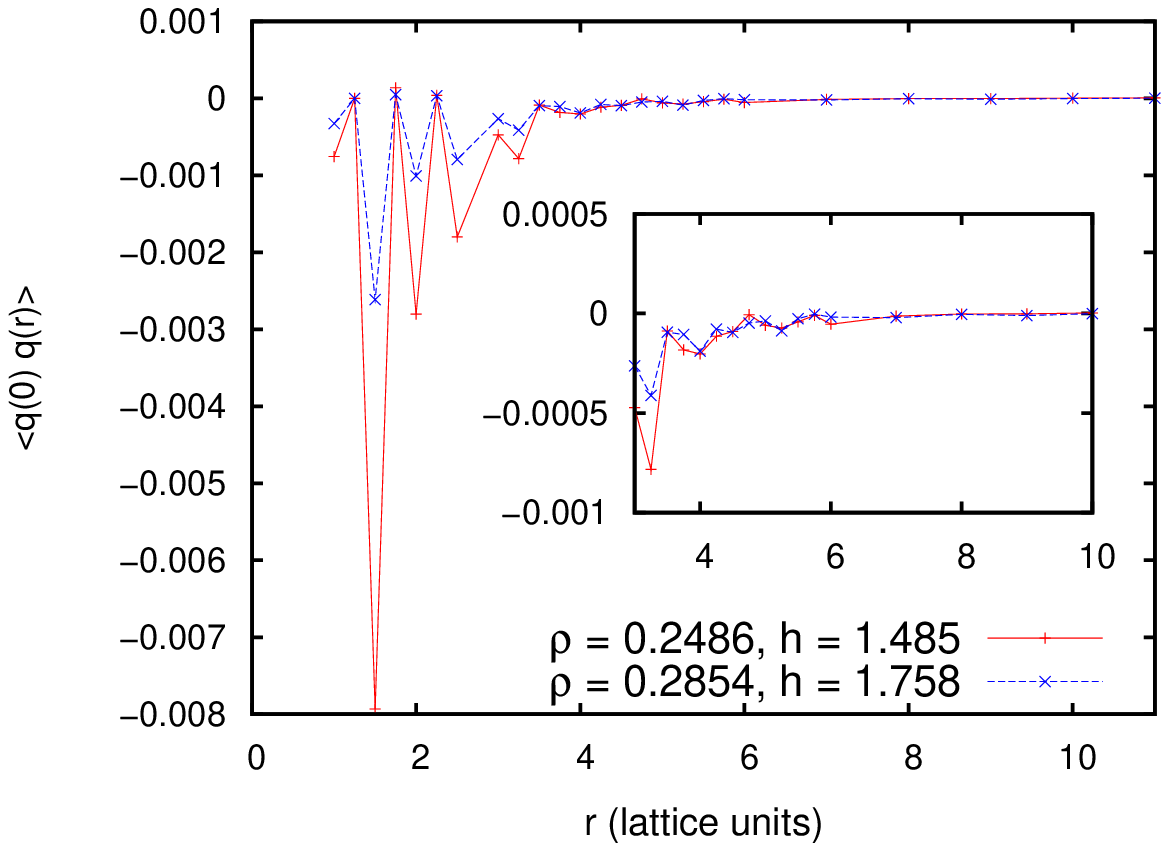,width=0.65\linewidth,height=0.4\textheight}
	 	\caption{\label{fig_simqq} MC : charge fluctuations' correlation function}
	\end{figure*}

\end{document}